\documentclass[twocolumn,pre,amsmath,amssymb]{revtex4}

\usepackage{amsmath}    % need for subequations
\usepackage{amsfonts}
\usepackage{amssymb}
\usepackage{graphicx}   % for figures
\usepackage{color}
\usepackage{dcolumn}
\usepackage{bm}

\begin{document}

\title{Modified jump conditions for parallel collisionless shocks}

\author{Antoine Bret}
\affiliation{ETSI Industriales, Universidad de Castilla-La Mancha, 13071 Ciudad Real, Spain}
 \affiliation{Instituto de Investigaciones Energ\'{e}ticas y Aplicaciones Industriales, Campus Universitario de Ciudad Real,  13071 Ciudad Real, Spain.}
 \email{antoineclaude.bret@uclm.es.}

\date{\today }

\begin{abstract}
Within the context of Magnetohydrodynamics (MHD), the properties of a parallel shock do not depend on the field strength, as the field and the fluid are disconnected for such a geometry. However, in the collisionless case, the field can sustain a stable anisotropy in the downstream, triggering a departure from the expected MHD behavior. In a recent work [A. Bret and R. Narayan, J. Plasma Phys. \textbf{84}, 905840604 (2018)], a theoretical model was presented allowing to derive the density ratio of a non-relativistic parallel collisionless shock in an electron/positron plasma, as a function of the field. Here we derive the entropy, pressure and temperature jumps stemming from this model. It is found to offer a transition between a 3D and a 1D downstream for the jumps in density, entropy, parallel temperature and parallel pressure.
\end{abstract}

\maketitle

\section{Introduction}
As fundamental processes in fluids, shockwaves have been under investigation since the 19th century \cite{Salas2007}.  In plasmas, the most developed theory of shockwaves has been elaborated within the context of Magnetohydrodynamics  (MHD) \cite{Kulsrud2005}. Yet, MHD relies on a number of assumptions (see below) that can be broken in collisionless plasmas, especially when magnetized \cite{BretApJ2020}.

It is therefore important to study departures from MHD in collisionless shocks physics. A variation of the density ratio $r$, for example, bears consequences on all the other ratios (temperature, pressure, etc) and even on the spectrum of accelerated particles since its index is proportional to $(r-1)^{-1}$ \cite{Blandford78}.

In this respect, parallel shocks, where the direction of propagation of the shock is parallel to the external magnetic field, are worthy of interest. Besides their abundance in space, at some location of a bow shock for example \cite{Treumann2009}, their MHD  properties are independent on the field strength \cite{Lichnerowicz1976,Kulsrud2005}. Therefore, any variation of such a shock when the field varies is necessarily a departure from its expected MHD behavior.

MHD assumes the pressure is isotropic both in the upstream and the downstream. While this assumption is justified in a collisional fluid, it is well-known that in the absence of collisions, a magnetic field can sustain a stable anisotropy \cite{Gary1993}. As a consequence, in case the upstream turns anisotropic as the plasma goes through the shock front, the resulting downstream anisotropy could be stable, triggering a departure from the MHD behavior \cite{BretJPP2017,BretJPP2018,BretApJ2020}.

In a recent article \cite{BretJPP2018}, a model was developed allowing to compute the downstream anisotropy in terms of the field strength and of the upstream properties. A field dependent expression of the parallel shock density ratio $r$ was derived, which, for a strong shock in a perfect fluid, goes from $r=4$ to $r=2$ in the strong field regime.

The model relies on the assumptions that the transit through the front is adiabatic in the perpendicular direction, but not in the parallel one. This can be understood picturing the fluid as being compressed between two converging plates parallel to the front, as it goes through it. The perpendicular temperatures therefore evolve according to the double adiabatic conditions given in Ref. \cite{CGL1956}. For the case of a parallel shock, this means they are simply conserved. As for the parallel temperature, it is given by the conservation equations \cite{BretJPP2018}.

The state of the downstream thus obtained is labelled ``Stage 1'' and is not isotropic. If the field is strong enough, Stage 1 can be stable. In such a case, it is the final state of the downstream. For weaker fields, Stage 1 can be firehose unstable \cite{BretJPP2018}. In that case, the plasma migrates to the firehose instability threshold, thus reaching the so-defined ``Stage 2'', stable by definition, and therefore final state of the downstream. Again the conservations equations, together with the marginal instability constraint, fully determine the properties of Stage 2. Recent Particle-In-Cell simulations are confirming the validity of this scenario \cite{Colby2021}.

In ref. \cite{BretJPP2018}, the aforementioned assumptions were used to derive the density jump of the shock. The goal of the present article is to derive, or further study, the full field dependent jump conditions for temperatures, pressures and entropy, and compare them to their MHD counterparts.

For completeness of the present work, some previous results are reminded in Section \ref{MHD}.

\section{Formalism and field dependent density ratio}\label{MHD}
We consider a non-relativistic shock in a pair plasma (electron/positron, or any 2 species of opposite charge and same mass). Such a system allows to deal with common parallel and perpendicular temperatures for the 2 species, whereas in a electron/ion plasma, we would have to account for different temperatures for the ions and the electrons \cite{Guo2017,Guo2018}. The shock propagates along the $x$ axis, which is also the direction of the field $\mathbf{B}_0$. Therefore, the $x$ direction is parallel to both the field and the flow. The $y$ and $z$ directions are normal to both the field and the flow.

The field strength is measured through the $\sigma$ parameter defined as,
\begin{equation}\label{eq:sigma}
\sigma = \frac{B_0^2/4\pi}{n_1v_1^2}.
\end{equation}

We assume an isotropic upstream. Using the subscripts ``1'' for the upstream quantities and ``2'' for the downstream, the matter, momentum and energy conservations equations read for an upstream adiabatic index $\gamma=5/3$,
\begin{eqnarray}
% \nonumber to remove numbering (before each equation)
  n_1v_1                                  &=& n_2v_2, \label{eq:conser1}\\
  n_1v_1^2 + P_1                          &=& n_2v_2^2 + P_{\parallel 2}, \label{eq:conser2}\\
  \frac{v_1^2}{2} + \frac{P_1}{n_1} + \underbrace{\frac{3}{2}\frac{P_1}{n_1}}_{U_1} &=&
                        \frac{v_2^2}{2} + \frac{P_{\parallel 2}}{n_2} + \underbrace{\frac{1}{2}\frac{P_{\parallel 2}+2P_{\perp 2}}{n_2}}_{U_2}.  \label{eq:conser3}
\end{eqnarray}
The momentum conservation equation (\ref{eq:conser2}) involves only the downstream parallel pressure $P_{\parallel 2}$ as it is the one producing the pressure force that pushes the plasma across the shock. On the same equation, the upstream pressure is simply $P_{\parallel 1}=P_1$. Likewise, the $P_i/n_i$ terms of the energy equation (\ref{eq:conser3}) read $P_1/n_1$ for the upstream and $P_{\parallel 2}/n_2$ for the downstream, since its physical origin can be traced back to the work of the pressure force (see ref. \cite{FeynmanVol2} \S 40-3).

Regarding the expression of the internal energy $U_i$, the upstream is assumed isotropic with $\gamma=5/3$, so that the internal energy term $U_1$ simply reads $3P_1/2n_1$. For the downstream, the perpendicular pressures are coupled via the Vlasov equation (see Ref. \cite{LandauKinetic} \S 53), so that the trace of the pressure tensor reads $P_{\parallel 2}+2P_{\perp 2}$. We do not assume any specific value of $\gamma$ for the downstream (see comments below Eq. \ref{eq:chi}).

For Stage 1, these equations are solved setting $P_{\perp 2}=n_2 k_BT_{\perp 2} = n_2 k_BT_1$ in Eq. (\ref{eq:conser3}). For Stage 2, $P_{\parallel 2}$ and $P_{\perp 2}$ are related to each other via the $B_0$ dependent firehose stability condition \cite{Gary1993,Gary2009},
\begin{equation}\label{eq:firehose}
A_2 \equiv \frac{T_{\perp 2}}{T_{\parallel 2}} = 1 - \frac{1}{\beta_{\parallel 2}},
\end{equation}
with,
\begin{equation}
\beta_{\parallel 2} = \frac{n_2k_BT_{\parallel 2}}{B_0^2/8\pi}.
\end{equation}

In the MHD approximation, considering downstream isotropy,  Eqs. (\ref{eq:conser1}-\ref{eq:conser3}) give the well-known jump conditions for the density, the pressure and the temperature \cite{Zeldovich},
\begin{eqnarray}\label{eq:MHD}
% \nonumber to remove numbering (before each equation)
  r \equiv \frac{n_2}{n_1} &=& \frac{\gamma +1}{\gamma -1}+\frac{2 \gamma }{(\gamma +1) \chi_1^2}, \\
  \frac{P_2}{P_1} &=& \frac{2 \chi_1^2}{\gamma +1}-\frac{\gamma -1}{\gamma +1}, \nonumber\\
  \frac{T_2}{T_1} &=& \frac{P_2/n_2}{P_1/n_1} = \frac{(\gamma -1) \chi_1^2 \left(-\gamma +2 \chi_1^2+1\right)}{2 (\gamma -1) \gamma +(\gamma +1)^2 \chi_1^2}, \nonumber
\end{eqnarray}
in terms of the parameter $\chi_1$,
\begin{equation}\label{eq:chi}
\chi_1 = \frac{v_1^2}{P_1/n_1}.
\end{equation}
Even though $\chi_1$ is close to the usual upstream Mach number, it is preferable to avoid using  a Mach number because we freeze some degrees of freedom of the plasma for Stage 1. As a consequence, a coherent treatment of Stage 1 and 2 is better achieved through this pseudo Mach number $\chi_1$. Indeed, we shall see in the sequel that the model offers a transition between a 3D and a 1D-like downstream, forbidding the definition of a Mach number with a fixed adiabatic index.

Following the aforementioned prescriptions for Stage 1 and 2, the following expression of the density ratio $r$ was derived in Ref. \cite{BretJPP2018},
\begin{equation}\label{eq:r}
r = \left\{
\begin{array}{r}
   \frac{5+\chi_1^2 \left(5-\sigma +\sqrt{\Delta}\right)}{2 \left(\chi_1^2+5\right)}, \sigma < \sigma_c, ~~~ \mathrm{(Stage~2)} \\
  \frac{2\chi_1^2}{\chi_1^2+3}, \sigma > \sigma_c, ~~~ \mathrm{(Stage~1)}
\end{array}
\right.
\end{equation}
where,
\begin{eqnarray}\label{eq:sigmac}
    \Delta &=& \frac{25}{\chi_1^4}-\frac{10 (\sigma+3)}{\chi_1^2}+(\sigma-9) (\sigma-1),  \nonumber \\
  \sigma_c &=& 1 -\frac{4}{\chi_1^2+3}-\frac{1}{\chi_1^2}.
\end{eqnarray}

\begin{figure}
\begin{center}
 \includegraphics[width=0.45\textwidth]{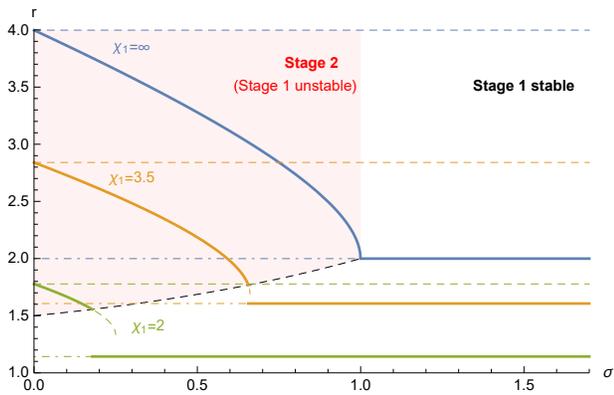}
\end{center}
\caption{Shock density ratio in terms of $\sigma$ for various values of $\chi_1$. At small $\sigma$, Stage 1 is unstable in the red shaded area and the system settles into Stage 2. At large $\sigma$, Stage 1 is stable and the downstream settles there. The thin horizontal dashed lines are the 1 and 3D MHD predictions for each $\chi$. The thin dashed colored curves show Stage 2 offers solutions even when Stage 1 is stable. They are mostly visible for $\chi_1=2$.}\label{fig:density}
\end{figure}

The density jump so defined is pictured on Figure \ref{fig:density} for various pseudo Mach numbers $\chi_1$. As will be checked in the sequel, Stage 1 is extremely anisotropic. It is therefore never stable for weak field. Hence, in this regime, the downstream systematically jumps to Stage 2. As the field increases, larger and larger anisotropies can be stabilized, until Stage 1 can be so. This is the reason why Stage 1 is the relevant state at high $\sigma$.

 Note that $\sigma_c$, the critical value of $\sigma$ at which the transition Stage 1 $\rightarrow$ Stage 2 occurs, is not the $\sigma_\Delta$ making $\Delta < 0$. Instead, it is the $\sigma$ below which Stage 1 is unstable because the field is too weak. Indeed, on Fig. \ref{fig:density}, the thin dashed colored curves show how  Stage 2 solutions slightly extend beyond $\sigma_c$, as $\sigma_\Delta > \sigma_c$. Yet, $\lim_{{\chi_1 \gg 1}} \sigma_\Delta = \sigma_c$ so that the difference is clearly visible for $\chi_1=2$, hardly visible for $\chi_1=3.5$, and invisible for $\chi_1=\infty$.

 For $\sigma=0$, Stage 2 density jump fits the 3D MHD jump (\ref{eq:MHD}) with $\gamma=5/3$. For $\sigma > \sigma_c$, Stage 1 density jump fits the 1D MHD jump (\ref{eq:MHD}) with $\gamma=3$. We see here that our model offers a transition between a 3D downstream for weak field, and a 1D downstream for strong field. These features will be retrieved with the entropy, the pressures and the temperatures jumps.

 \begin{figure}
\begin{center}
 \textbf{(a)}\\\includegraphics[width=0.45 \textwidth]{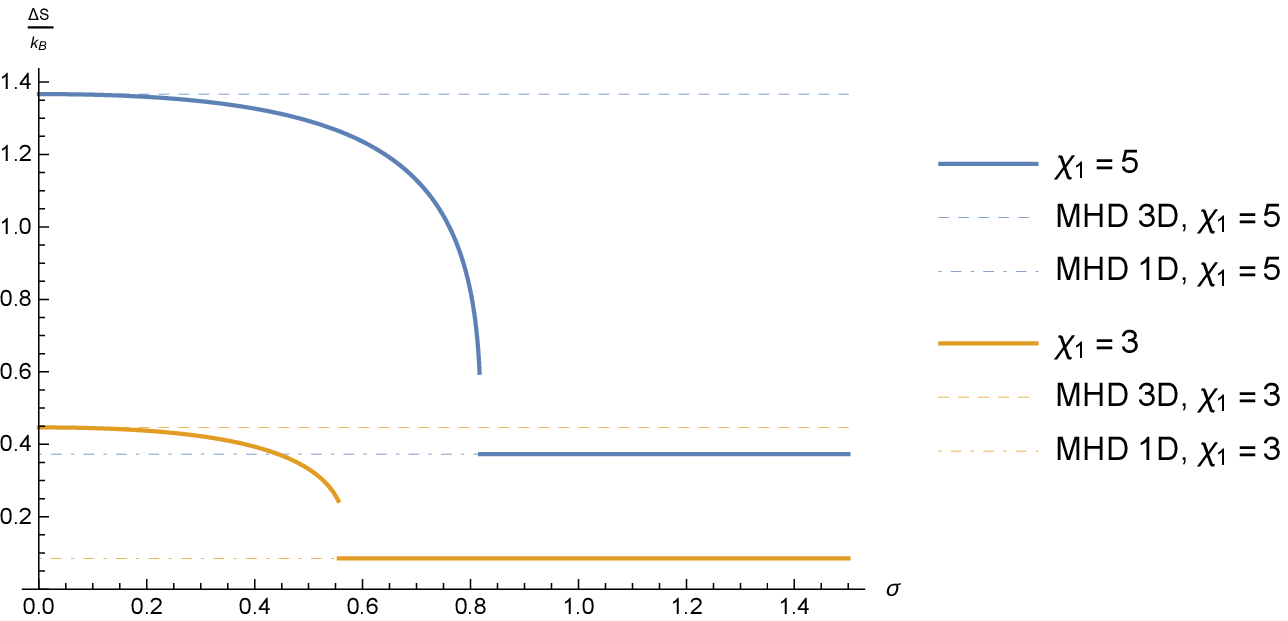} \\ \textbf{(b)}\\\includegraphics[width=0.4 \textwidth]{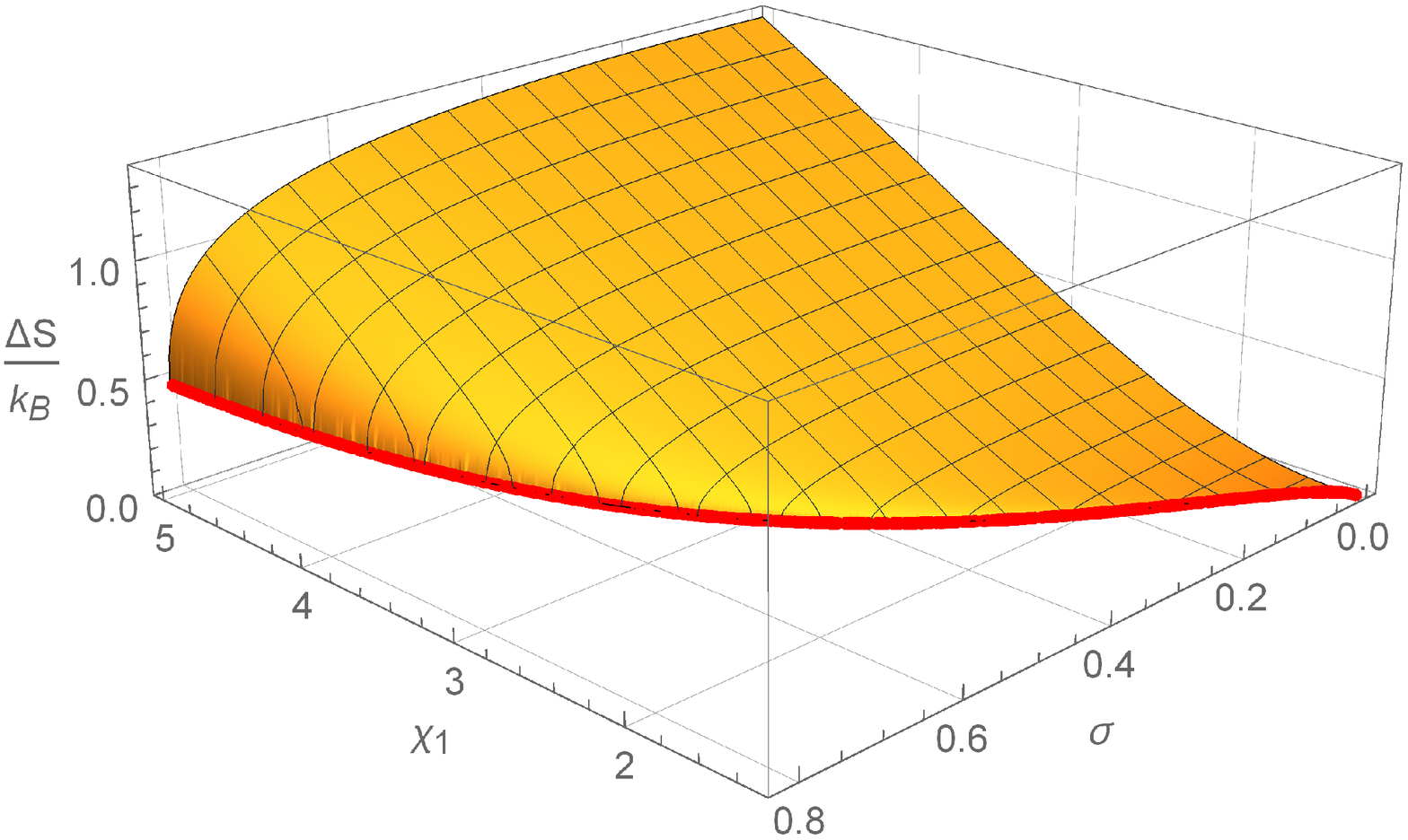}
\end{center}
\caption{(a): Entropy jump in terms of $\sigma$ for various values of $\chi_1$. (b): Entropy jump for Stage 2 from Eq. (\ref{eq:Ds2}). The red line pertains to $\Delta=0$, where $\Delta=0$ is given by Eq. (\ref{eq:sigmac}).}\label{fig:entropy}
\end{figure}

\section{Entropy jump}
Since the calculation of the entropy allows to complete the physical meaning of the pseudo Mach number $\chi_1$, we now present its evaluation. The entropy jump was computed in Ref. \cite{BretJPP2018} for Stage 1 and Stage 2 and is here further studied. It reads,

\begin{itemize}
  \item For Stage 1,
  \begin{equation}\label{eq:Ds1}
\Delta s_1 = \frac{1}{2}k_B\ln \left(\frac{\left(\chi_1^2-1\right) \left(\chi_1^2+3\right)^3}{16 \chi_1^6}\right).
\end{equation}
  \item For Stage 2,
\begin{equation}\label{eq:Ds2}
\Delta s_2  = \frac{1}{2}k_B\ln \left[   \frac{A_2^2}{r^5} \left(  \frac{(r-1) \chi_1^2}{r}+1 \right)^3 \right],
\end{equation}
\end{itemize}
where $k_B$ is the Boltzmann constant and $A_2$ the downstream anisotropy (see Eq. \ref{eq:firehose}). The resulting entropy jump is plotted on Fig. \ref{fig:entropy}(a)  in terms of $\sigma$ for various values of $\chi_1$.
According to Eq. (\ref{eq:Ds1}), Stage 1 reaches $\Delta s_1=0$ for $\chi_1=\sqrt{3}$. Regarding Stage 2, its entropy jump $\Delta s_2$ can be studied from Eq. (\ref{eq:Ds2}). The result is plotted on Fig. \ref{fig:entropy}(b), with the thick red line picturing the frontier $\Delta=0$. Some algebra shows that $\Delta s_2$  reaches 0 for $(\chi_1,\sigma)=(\sqrt{5/3},0)$, with $\Delta s_2(\chi_1<\sqrt{5/3},\sigma)<0$, or undefined because of $\Delta < 0$.

Therefore, if the model is to offer a physically meaningful solution, it requires $\chi_1 > \sqrt{3}$ because of Stage 1, and $\chi_1>\sqrt{5/3}$ because of Stage 2, that is, $\chi_1 > \max(\sqrt{3},\sqrt{5/3})=\sqrt{3}$.

It is interesting to compare these calculations with the entropy jump in the MHD approximation in 1D and 2D. To do so we assume isotropic distributions in the upstream and the downstream, of the form,
\begin{equation}\label{eq:fi}
 f_i= \frac{n_i}{(2\pi k_BT_i/m)^{D/2}}\exp \left(-\frac{m v^2}{2 k_BT_i}\right),
\end{equation}
where the subscript $i$ refers to the upstream or the downstream ($i=1$ and 2 respectively), and $D$ is the dimension. The entropy can be computed as $S=-\int f_i \ln f_id^Dv$ \cite{LandauStat} and the entropy jump per particle reads,
\begin{equation}\label{eq:Deltas}
\Delta s = \frac{S_2}{n_2} - \frac{S_1}{n_1} = \frac{D}{2} \ln \left(  \frac{T_2}{T_1}  \right)-\ln \left(  \frac{n_2}{n_1}  \right).
\end{equation}

Figure \ref{fig:entropy}(a) features the value of the ``MHD entropy'' (\ref{eq:Deltas}) for $D=1$ and 3. For $\sigma=0$, Stage 2 is 3D isotropic and the MHD entropy jump for $D=3$ is recovered. For $\sigma > \sigma_c$, the entropy jump of Stage 1 fits the MHD entropy jump for $D=1$. Hence, the model switches between a 3D and a 1D downstream.

\section{Pressures jump}
Since the downstream is no longer considered isotropic like in MHD, we have to consider 2 pressure jumps instead of 1. The downstream pressures can be computed from Eqs. (\ref{eq:conser1}-\ref{eq:conser3}) according to the following procedure. First, use the matter conservation equation (\ref{eq:conser1}) to eliminate $v_2$ from the momentum and energy equations (\ref{eq:conser2},\ref{eq:conser3}). The new Eq. (\ref{eq:conser2}) can be used to express $P_{\parallel 2}$. In the same way, the new Eq. (\ref{eq:conser3}) can also be used to express $P_{\parallel 2}$. Note that this later expression is different whether we consider Stage 1 or Stage 2.
\begin{itemize}
  \item For Stage 1, where $T_\perp$ is conserved, we set $P_{\perp 2}= n_2 k_BT_{\perp 2} = (n_2/n_1) n_1k_BT_1 = rP_1$.
  \item For Stage 2, we set $P_{\perp 2} = P_{\parallel 2}(1-\beta_{\parallel 2}^{-1})$ in order to fulfil the firehose stability criteria (\ref{eq:firehose}).
\end{itemize}
The 2 resulting expressions for $P_{\parallel 2}$ can be equalled to derive Eq. (\ref{eq:r}) for $n_2/n_1$. In turn, Eq. (\ref{eq:r}) can then be used to express $P_{\parallel 2}$, and then $P_{\perp 2}$. The results are,
\begin{itemize}
  \item For Stage 1,
  \begin{eqnarray}\label{eq:DP1}
 % \nonumber to remove numbering (before each equation)
   P_{\parallel 2}  &=&   \frac{1}{2} \left(n_1 V_1^2-P_1\right) = \frac{1}{2}\left(\chi_1^2-1\right) P_1 ,\\
   P_{\perp 2} &=& n_2 k_B T_{\perp 2} = n_2 k_B T_{\perp 1} = n_2 \frac{P_1}{n_1} \nonumber \\
              &=& \frac{2 \chi_1^2}{\chi_1^2+3}P_1,
 \end{eqnarray}
  so that the downstream anisotropy $A_2$ for Stage 1 reads,
\begin{equation}\label{eq:A2P1}
  A_2=\frac{P_{\perp 2}}{P_{\parallel 2}} = \frac{4 \chi_1^2}{(\chi_1^2+3)(\chi_1^2-1)}.
\end{equation}
  \item For Stage 2,
  \begin{eqnarray}\label{eq:DPS2}
P_{\parallel 2} &=& P_1 \left(\frac{(r-1) \chi_1^2}{r}+1\right) ,  \\
P_{\perp 2} &=& \frac{P_{\perp 2}}{P_{\parallel 2}}P_{\parallel 2} =  A_2 P_{\parallel 2},
\end{eqnarray}
with,
\begin{equation}\label{eq:A2Stage2}
 A_2=\frac{T_{\perp 2}}{T_{\parallel 2}}=\frac{(r-2) (r-1) \chi_1^2+r (5 r-3)}{2( r + (r-1) \chi_1^2  )}.
\end{equation}
\end{itemize}

\begin{figure}
\begin{center}
 \includegraphics[width=0.45\textwidth]{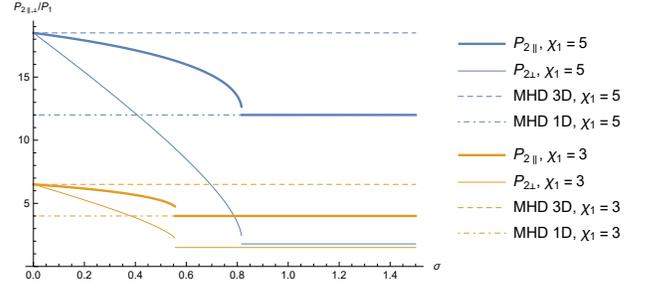}
\end{center}
\caption{Downstream parallel (thick) and perpendicular (thin) pressures in terms of $\sigma$ for various values of $\chi_1$. The dashed lines pictures the MHD results.}\label{fig:pressure}
\end{figure}

Figure \ref{fig:pressure} pictures the downstream parallel and perpendicular  pressures in terms of $\sigma$ for various values of $\chi_1$. Similarly to the density and entropy plots, pressures pertain to Stage 2 at low $\sigma$, and to Stage 1 at high $\sigma$, when the field is strong enough to stabilize it. The transition occurs for $\sigma=\sigma_c$ given by Eq. (\ref{eq:sigmac}), below which Stage 1 is unstable.

In units of $P_1$, the downstream perpendicular pressure $P_{\perp 2}$ systematically comes close to $P_1$ for strong field since Stage 1 has $T_{\perp 2}=T_1$ and a density ratio of order unity.

Figure \ref{fig:pressure} also compares the pressure jumps given by the model with the MHD ones. For $\sigma=0$ the model fits exactly the 3D MHD jump. For strong field, namely $\sigma > \sigma_c$, the parallel pressure fits the 1D MHD jump.

For low $\sigma$'s, the following expressions can be derived expanding Stage 2 pressures to first order in $\sigma$,
\begin{eqnarray}
P_{\parallel 2} &=& P_{2,MHD}\left( 1-\frac{ \chi_1^2 (\chi_1^2+5)}{9 (\chi_1^2-2) \chi_1^2+5}\sigma  \right), ~~ \sigma \ll \sigma_c  \\
P_{\perp 2} &=& P_{2,MHD}\left( 1-  \frac{\chi_1^2 (7 \chi_1^2-5)}{9 (\chi_1^2-2) \chi_1^2+5} \sigma  \right), ~~ \sigma \ll \sigma_c
\end{eqnarray}
and
\begin{equation}
 A_2 = 1 - \frac{2  \chi_1^2}{3 \chi_1^2-1}\sigma, ~~ \sigma \ll \sigma_c.
\end{equation}
In the strong shock limit $\chi_1 \gg 1$, we would then have,
\begin{eqnarray}
P_{\parallel 2} &=& P_{2,MHD}\left( 1-\frac{1}{9}\sigma  \right), ~~ \sigma \ll \sigma_c,~\chi_1 \gg 1   \\
P_{\perp 2} &=& P_{2,MHD}\left( 1-  \frac{7}{9} \sigma  \right), ~~ \sigma \ll \sigma_c,~ \chi_1 \gg 1  \\
 A_2 &=& 1 - \frac{2}{3}\sigma, ~~ \sigma \ll \sigma_c,~ \chi_1 \gg 1.
\end{eqnarray}

In the opposite limit $\sigma > \sigma_c$, Stage 1 is to be considered. The parallel and perpendicular pressures, together with the anisotropy, are therefore given by Eqs. (\ref{eq:DP1},\ref{eq:A2P1}).

\begin{figure}
\begin{center}
 \includegraphics[width=0.45\textwidth]{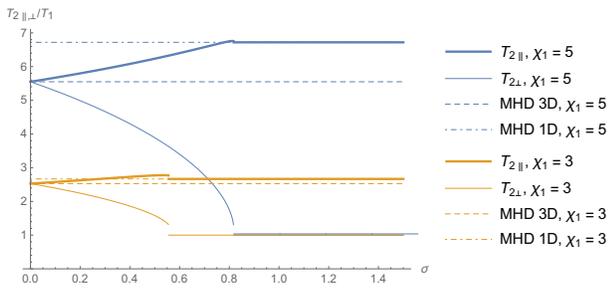}
\end{center}
\caption{Downstream parallel (thick) and perpendicular (thin) temperatures in terms of $\sigma$ for various values of $\chi_1$. The dashed lines pictures the MHD results. The thin blue horizontal line pertaining to $T_{\perp 2}$ for $\chi_1=5$ and $\sigma > 0.8$  has been slightly shifted up to be visible, as it exactly overlaps with the orange one.}\label{fig:temp}
\end{figure}

\section{Temperatures jump}
Here again we need to consider 2 temperature jumps instead of 1. From $P_{2\parallel,\perp}=n_2k_BT_{2\parallel,\perp}$ we simply derive $T_{2\parallel,\perp}=P_{2\parallel,\perp}/n_2k_B$. The algebra gives,
\begin{itemize}
  \item For Stage 1,
  \begin{eqnarray}\label{eq:DT1}
 % \nonumber to remove numbering (before each equation)
   T_{\perp 2} &=& T_1, \\
   T_{\parallel 2}  &=&   T_1 \frac{ \left(\chi_1^2-1\right) \left(\chi_1^2+3\right)}{4 \chi_1^2}.
 \end{eqnarray}
  \item For Stage 2,
  \begin{eqnarray}\label{eq:DT2}
 % \nonumber to remove numbering (before each equation)
   T_{\parallel 2}  &=& T_1\frac{\left(\chi_1^2+5\right) \left(\chi_1^2 \left(\sqrt{\Delta }+s+3\right)+3\right)}{4 \chi_1^2 \left(\sqrt{\Delta }-s+5\right)+20},  \\
   T_{\perp 2} &=&  T_1\frac{\left(\chi_1^2+5\right) \left(\chi_1^2 \left(\sqrt{\Delta }-3 s+3\right)+3\right)}{4 \chi_1^2 \left(\sqrt{\Delta }-s+5\right)+20},
 \end{eqnarray}
  where $\Delta$ is given by Eq. (\ref{eq:sigmac}).
\end{itemize}

Figure \ref{fig:temp} pictures the downstream parallel and perpendicular temperatures in terms of $\sigma$ for various values of $\chi_1$. As was the case for the entropy, the density and the pressures, for $\sigma=0$ the temperature jumps are given by their 3D MHD counterpart. Regarding the strong field regime $\sigma > \sigma_c$, the parallel temperature jump is given by the 1D MHD result, as was the case for the parallel pressure.

At large $\sigma$, that is, $\sigma > \sigma_c$, the perpendicular temperature goes to $T_1$ by definition of Stage 1. Therefore, in units of $T_1$, all curves on Fig. \ref{fig:temp}  settle to unity at high $\sigma$ regardless of $\chi_1$. For the Stage 1 thin horizontal lines pertaining to $\chi_1=5$ and $3$ to be visible on Fig. \ref{fig:temp}, the blue one has been slightly shifted up.

For low $\sigma$'s, the following Stage 2 expressions can be derived,
\begin{eqnarray}
T_{\parallel 2} &=& T_{2,MHD}\left( 1 + \frac{2 \chi_1^2 \left(\chi_1^2-3\right)}{9 \left(\chi_1^2-2\right) \chi_1^2+5}   \sigma  \right), ~~ \sigma \ll \sigma_c,  \\
T_{\perp 2} &=& T_{2,MHD}\left( 1-   \frac{4 \chi_1^2 \left(\chi_1^2-1\right)}{9 \left(\chi_1^2-2\right) \chi_1^2+5}  \sigma  \right), ~~ \sigma \ll \sigma_c,
\end{eqnarray}
with the strong shock limits,
\begin{eqnarray}
T_{\parallel 2} &=& T_{2,MHD}\left( 1 + \frac{2}{9}  \sigma  \right), ~~ \sigma \ll \sigma_c,~ \chi_1 \gg 1,  \\
T_{\perp 2} &=& T_{2,MHD}\left( 1-  \frac{4}{9}  \sigma  \right), ~~ \sigma \ll \sigma_c,~ \chi_1 \gg 1.
\end{eqnarray}

In the high field regime, namely $\sigma > \sigma_c$, Eqs. (\ref{eq:DT1}) give the corresponding temperatures.

\section{Conclusion}
We have systematically derived the entropy, pressure and temperature jumps arising from the model described in Ref. \cite{BretJPP2018} which aims at modelling the kinetic effects modifying the properties of a parallel collisionless shock. The results apply to pair plasmas. According to MHD theory, these properties should not vary with the strength of the field. Yet, kinetic effects do modify them by allowing for a stable downstream anisotropy in the presence of a magnetic field.

The downstream anisotropy is generated as the plasma goes through the shock front. In that process, the perpendicular temperatures evolve adiabatically while the heating, i.e., the entropy increase, goes into the direction parallel to the flow. For the present case of a parallel shock, the adiabatic evolution of the perpendicular temperatures simply means that they are conserved. The resulting downstream state is labelled Stage 1 and can be stable or unstable. For too weak a field, it is firehose unstable and migrates to Stage 2, which is by definition marginally firehose stable. In that case, the end state of the downstream is therefore Stage 2. For strong enough a field, Stage 1 is stable and is therefore the end state of the downstream.

For $\sigma=0$, Stage 2 imposes downstream isotropy so that the 3D MHD results are recovered for the jumps of density, entropy, pressure and temperature. At high sigma, the jumps in density, entropy, parallel pressure and temperatures, all fit the 1D MHD predictions. Hence, the model here describes a transition from a 3D to a 1D downstream.

Applying these results to electrons/ions shocks seems premature. As already alluded in Section \ref{MHD}, temperature differences between ions and electrons could affect the proposed scenario. To our knowledge, no simulation studies have been conducted so far for parallel electrons/ions shocks at $\sigma > 1$, where notable departure from the MHD jump conditions on a macro-scale is expected. At lower $\sigma$'s, no significant departure from the MHD density ratio has been noticed so far in the literature \cite{BretApJ2020}. Shocks in electrons/ions  will therefore be the topic of forthcoming theoretical and numerical works.

\section{Acknowledgments}
A.B. acknowledges support by grants ENE2016-75703-R from the Spanish
Ministerio de Ciencia, Innovaci\'{o}n y Universidades and SBPLY/17/180501/000264 from the Junta de
Comunidades de Castilla-La Mancha.

\section{DATA AVAILABILITY}
The data that support the findings of this study are available from the corresponding author upon reasonable request.

%\bibliographystyle{unsrt}
%\bibliography{BibBret}

\end{document}